\documentclass[sigconf]{acmart}

\AtBeginDocument{%
  \providecommand\BibTeX{{%
    Bib\TeX}}}

\setcopyright{acmlicensed}
\copyrightyear{2026}
\acmYear{2026}
\setcopyright{cc}
\setcctype{by}
\acmConference[FORGE '26]{2026 IEEE/ACM Third International Conference on AI Foundation Models and Software Engineering}{April 12--13, 2026}{Rio de Janeiro, Brazil}
\acmBooktitle{2026 IEEE/ACM Third International Conference on AI Foundation Models and Software Engineering (FORGE '26), April 12--13, 2026, Rio de Janeiro, Brazil}
\acmPrice{}
\acmDOI{10.1145/3793655.3793731}
\acmISBN{979-8-4007-2477-0/2026/04}

\usepackage[utf8]{inputenc}
\usepackage{xspace}
\usepackage{multirow}
\usepackage{subfig} 
\usepackage{balance} 
\usepackage{rotating}
\usepackage{enumitem}
\usepackage{listings}
\usepackage{lscape}

\definecolor{codegreen}{rgb}{0,0.6,0}
\definecolor{codegray}{rgb}{0.5,0.5,0.5}
\definecolor{codepurple}{rgb}{0.58,0,0.82}
\definecolor{backcolour}{rgb}{0.95,0.95,0.92}
\definecolor{medium-blue}{rgb}{0,0,1}
\definecolor{darkblue}{HTML}{003566}
\definecolor{lightgray}{gray}{0.97}

\hypersetup{
  colorlinks=true,
  linkcolor=black,
  urlcolor=medium-blue,
  citecolor=black
}

\lstdefinestyle{custompython}{
  backgroundcolor=\color{backcolour},
  commentstyle=\color{codegreen},
  keywordstyle=\color{magenta},
  numberstyle=\tiny\color{codegray},
  stringstyle=\color{codepurple},
  basicstyle=\scriptsize\ttfamily,
  breaklines=true,
  captionpos=b,
  numbers=left,
  numbersep=5pt,
  showstringspaces=false,
  tabsize=2
}
\newcommand{\name}{\textit{AutoReSpec}\xspace}
\newcommand{\leaderboardBench}{\textit{\name{}Bench}\xspace}
\newcommand{\llmrecommender}{\texttt{LLM recommender}\xspace}
\newcommand{\promptgenerator}{\texttt{Prompt generator}\xspace}

\newcounter{NumObservations}
\newcommand{\finding}[2][]{%
  \par\noindent
  \begingroup
  \setlength{\fboxsep}{6pt}%
  \colorbox{gray!10}{%
    \parbox{\dimexpr\linewidth-2\fboxsep\relax}{%
      \textbf{#1\ifx&#1&Finding~\arabic{NumObservations}\fi:}~#2%
    }%
  }%
  \endgroup
  \par
  \stepcounter{NumObservations}%
}

\usepackage{algorithmic}
\usepackage{graphicx}
\usepackage{textcomp}
\usepackage{graphicx}
\usepackage{subcaption}
\usepackage{adjustbox}
\usepackage{multirow}

\usepackage{tabularx}
\def\BibTeX{{\rm B\kern-.05em{\sc i\kern-.025em b}\kern-.08em
    T\kern-.1667em\lower.7ex\hbox{E}\kern-.125emX}}

\usepackage[ruled,linesnumbered]{algorithm2e}

\begin{document}

\title{\name: A Framework for Generating Specification using Large Language Models}

\author{Ragib Shahariar Ayon}
\affiliation{%
  \institution{Texas State University}
  \city{San Marcos}
  \state{TX}
  \country{USA}
}
\email{ipd21@txstate.edu}
\orcid{0009-0006-6372-5000}

\author{Shibbir Ahmed}
\affiliation{%
  \institution{Texas State University}
  \city{San Marcos}
  \state{TX}
  \country{USA}
}
\email{shibbir@txstate.edu}
\orcid{0000-0003-1183-883X}

\begin{abstract}
\label{sec:abstract}
Formal specification generation has recently drawn attention in software engineering as a way to improve program correctness without requiring manual annotations. Large Language Models (LLMs) have shown promise in this area, but early results reveal several limitations. Generated specifications often fail verification due to syntax errors, logical inaccuracies, or incomplete reasoning, especially in programs with loops or branching logic. Techniques like SpecGen and FormalBench attempt to address this through prompting and benchmarking, but they typically rely on static prompts and do not offer mechanisms for recovering from failure or adapting to different program structures. In this paper, we present \name, a collaborative framework that combines open and closed-source LLMs for verifiable specification generation. \name dynamically chooses an LLM pair and prompt configuration based on the structure of the input program. If the primary LLM fails to produce a valid output, a collaborative model is invoked, using validator feedback to refine and correct the specification. This two-stage design enables both speed and robustness. We evaluate \name on a new benchmark of 72 real-world and synthetic Java programs. Our results show that it achieves 67 passes out of 72, outperforming SpecGen and FormalBench in both Success Probability and Completeness. Our experimental evaluation achieves a 58.2\% success probability and a 69.2\% completeness score, while cutting evaluation time by 26.89\% on average compared to prior methods. Together, these results demonstrate that AutoReSpec offers a scalable, efficient, and reliable approach to LLM-based formal specification generation.
\end{abstract}


\maketitle



\section{Introduction}
\label{sec:introduction}
\begin{figure*}[!t]
\centering
	\includegraphics[width=\linewidth]{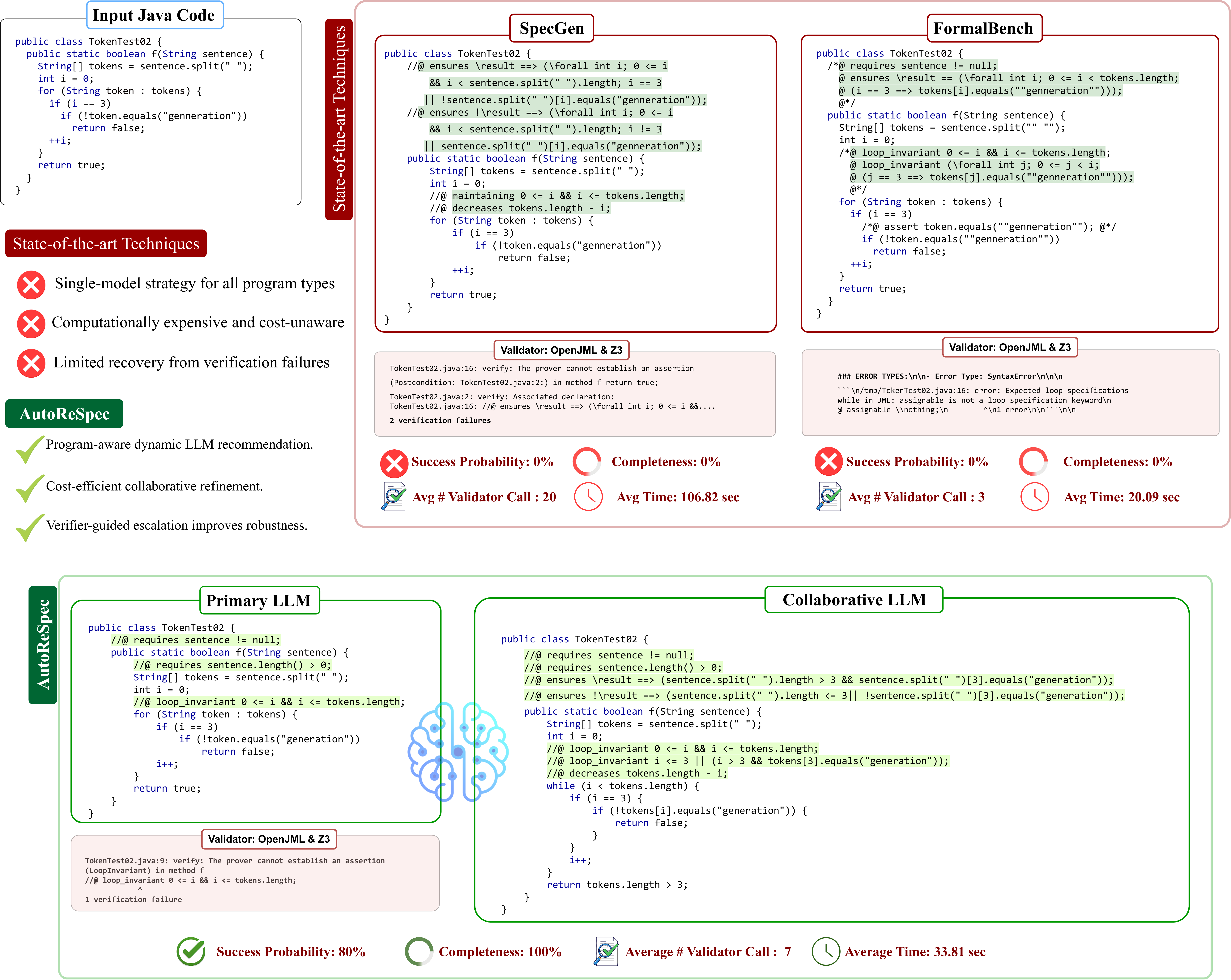}
    \Description{A three-column comparison for the Java method \texttt{TokenTest02} showing outputs from SpecGen, FormalBench, and \name. The SpecGen and FormalBench columns include candidate JML specifications that fail OpenJML verification, while the \name column shows a corrected JML specification that verifies successfully after collaborative refinement with verifier feedback, with accompanying summary metrics for success, completeness, validator calls, and runtime.}
	\footnotesize
	\caption{Motivating example illustrating Prior tools like SpecGen and FormalBench fail on the Java method \texttt{TokenTest02}, achieving 0\% success probability and completeness under OpenJML with Z3. In contrast, \name’s collaborative pipeline combining an open-source model with a proprietary fallback yields a correct, verifiable specification with 100\% success and completeness, fewer validator calls, and faster runtime.}
	\label{fig:motiv}
\end{figure*}
Formal specifications play a foundational role in reasoning about program behavior. They enable a range of tasks in software engineering, including automated verification~\cite{flanagan2002jml}, test generation~\cite{ernst2007daikon}, and program synthesis~\cite{alur2013syntax}. Despite their theoretical value, formal specifications are largely absent in real-world codebases. Writing them by hand requires deep expertise in formal methods, familiarity with specification languages like JML~\cite{leavens2006jml}, and significant time investment, especially for programs with complex control flow or semantic constraints. Several works have explored ways to automate this process~\cite{flanagan2001houdini, ernst2007daikon, molina2021evospex, wen2024enchanting}, with recent work focusing on leveraging LLMs to learn formal specifications directly from code.

A number of LLM-based approaches have been proposed for this task. SpecGen~\cite{specgen} applies a few-shot conversational prompting and mutation-based specification generation to improve verification success. Pei et al.~\cite{pei2023can} investigate whether LLMs can synthesize inductive invariants with fine-tuned models. FormalBench~\cite{formalbench} evaluates LLMs on a broader set of semantic reasoning tasks, using specification synthesis as a proxy for deeper program understanding. While these efforts demonstrate early promise, they face important limitations. SpecGen and Pei et al. lack generality across model families and fail to recover when initial specifications are unverifiable. FormalBench, by design, focuses on evaluation rather than generation, and does not investigate interactive or collaborative refinement workflows. None of these systems supports adaptive prompting or dynamic model selection tailored to program complexity.

To address these challenges, we introduce \name, a collaborative LLM framework for synthesizing formal specifications in Java Modeling Language (JML). Although \name\ is conceptually language-agnostic, supporting any setting where code and specification pairs can be verified, our implementation targets Java and JML due to the availability of mature verification tooling and benchmark datasets. \name is built on three key ideas: (1) dynamically selecting the optimal LLM pair (primary and fallback) and few-shot configuration based on the program type; (2) generating and refining prompts through a conversational loop guided by validator feedback; and (3) selectively invoking a collaborative model upon primary-model failure. \name first runs the primary LLM in a validator-guided refinement loop; if it still fails within the refinement budget, the last candidate specification and verifier error are forwarded to the collaborative LLM for focused recovery.

Developing such an end-to-end system poses several technical challenges. First, different programs require different prompt strategies and model capabilities. Second, verification failures require error-guided prompt refinement that is both specific enough to resolve issues and general enough to generalize across models. Third, integrating multiple LLMs and verifier iterations without excessive latency or token overflow requires careful prompt truncation, memory management, and recovery logic. \name mitigates these challenges through an adaptive recommender for model selection, a validator-guided refinement engine for error correction, an efficient prompt management module for context control, and a collaborative fallback mechanism for recovery.

Under the dynamic model selection configuration (RQ1), the collaborative version of \name\ verified 119 out of 120 programs from SpecGenBench, achieving 69.32\% success probability (SP) and 60.33 \% completeness (C).
On the full 72-program benchmark (RQ2) combining SpecGenBench, SV-COMP, and real-world OpenJML issues, \name\ achieved 58.2\% SP and 69.2\% C, while reducing average generation time by 26.89\% compared with prior tools. Validator logs further confirmed stable recovery behavior across diverse program types.

\noindent Our contributions are as follows:

\begin{itemize}
    \item We propose \name, a novel collaborative framework that recommends LLM-model pairs based on program structure and uses verifier-guided conversational refinement to generate verifiable specifications.
    
    \item We design a collaborative prompt engineering strategy that combines few-shot examples, validator feedback, and automated prompt truncation to support long-context interactions and enable targeted error resolution.
    
    \item We utilize a mutation-based completeness score, success probability, and number of passes, evaluation time to rank the LLMs for recommendation.
    
    \item We construct \leaderboardBench, a new benchmark combining real-world GitHub issues, loop-heavy SV-COMP programs, and challenging cases from SpecGenBench, and release a public \textbf{leaderboard} for reproducible evaluation~\cite{AutoReSpecLeaderBoard}.
    
    \item We release our full framework~\cite{AutoReSpecReplication}, including code, dataset, and VS Code extension~\cite{AutoReSpecVSCodeExt}, as an open-source tool chain to support future work in specification generation and LLM-based reasoning.
\end{itemize}

\noindent The remainder of this paper is organized as follows. Section~\ref{sec:motivation} presents a motivating example that highlights the key challenges in automated specification generation. Section~\ref{sec:framework} details the design of the \name\ framework. Section~\ref{sec:experimantaldesign} describes the experimental setup, while Section~\ref{sec:results-evaluation} reports the results and analysis. Section~\ref{sec:related-works} reviews related research, and Section~\ref{sec:threats} discusses threats to validity. Finally, Section~\ref{sec:conclusion} concludes the paper.
\section{Motivation}
\label{sec:motivation}
Automatically generating formal specifications is challenging because formal languages like JML are highly sensitive to syntax and require precise logical consistency. Even small mistakes, such as a misplaced quantifier, an incorrect assignable clause, or a wrong logical operator, can make an otherwise sound specification unverifiable. The challenge becomes greater in real-world Java programs that contain complex control flow and subtle loop invariants, which are often difficult to capture correctly in formal terms.

Figure~\ref{fig:motiv} illustrates these challenges using the \texttt{TokenTest02} method, which processes a string of tokens. The goal is to produce valid JML specifications. We compare three setups: SpecGen~\cite{specgen}, FormalBench~\cite{formalbench}, and our proposed \name. SpecGen uses few-shot prompting combined with mutation-based validation to iteratively improve JML specifications through verifier feedback. In contrast, FormalBench~\cite{formalbench} provides a standardized evaluation framework that benchmarks LLM-generated specifications using fixed prompts and evaluates consistency, completeness, and robustness under semantics-preserving transformations. Following FormalBench, a state-of-the-art (SOTA) benchmark setup, we adopt its specification-generation configuration as one of our baseline configurations, as it frames specification synthesis as a semantic-reasoning task and provides calibrated prompts and verification infrastructure suitable for fair comparison.

We evaluated each approach over ten independent trials to calculate the success probability ($SP$) and completeness ($\mathcal{C}$), averaging results across runs to account for stochastic model behavior. $SP$ measures the fraction of trials that produce verifiable specifications for a program, while $\mathcal{C}$ represents the average ratio of killed to total mutants, indicating fault-detection coverage. Both baseline systems fail in this example: SpecGen generates JML syntax errors, and FormalBench produces semantically inconsistent specifications that fail verification. Both SOTA methods use a single-model strategy across all program types, which does not account for program complexity or costs. This example also reflects two broader observations that motivate our design. First, LLM performance is program-type-dependent: models that work well on simple programs can fail on programs with multi-path control flow, which require stronger reasoning about invariants. Second, relying on a single model for all programs is often cost-inefficient, while relying only on smaller models can reduce verification success on harder cases. These observations motivate a dynamic LLM recommender that selects models based on program type and a collaborative strategy that uses a cost-efficient primary model with a stronger model invoked only when needed.

In contrast, \name identifies the program as a multi-path loop and uses its LLM recommender to select Llama~3 as the primary LLM and GPT-4o as the collaborative LLM. It generates an initial candidate and, if verification fails, forwards the last candidate and validator feedback to the collaborative LLM for recovery. This collaborative step corrects the specification and produces a valid JML-annotated method, achieving 100\% $SP$ and $\mathcal{C}$ with fewer validator calls and lower runtime.

Taken together, the example yields three key takeaways: (1) naive prompting is insufficient for verifiable specification generation, even for simple programs; (2) recovery from validation failure requires guided refinement and verifier feedback; and (3) coordinated model collaboration improves accuracy, efficiency, and robustness. These insights directly motivate the design of \name, presented in the following section.
\section{Framework}
\label{sec:framework}
\subsection{Overview}
\label{subsec:overview}

\begin{figure*}[!t]
\centering
\includegraphics[width=\linewidth,trim={0cm 0cm 0cm 0cm},clip]{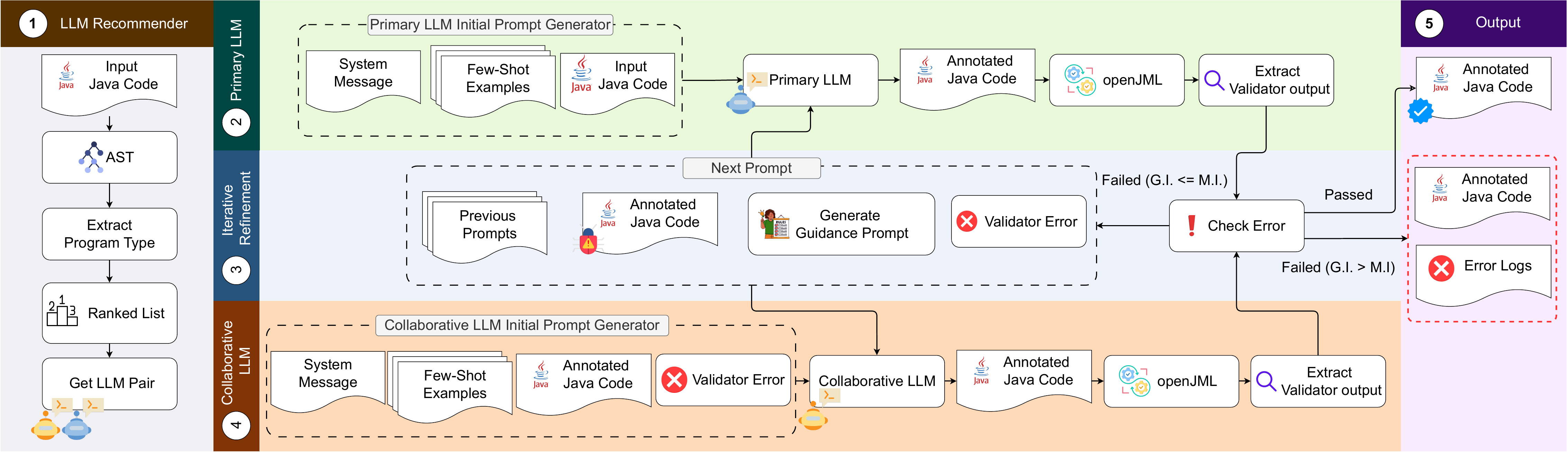}
\Description{Block diagram of the \name framework showing the end-to-end pipeline: an input Java program is analyzed and routed by an LLM recommender to select a model pair and prompt configuration; the primary LLM generates candidate JML specifications that are checked by OpenJML; verifier feedback drives iterative refinement; if the primary model fails within the budget, a collaborative LLM is invoked to revise the last candidate using the error trace; the process outputs a verifiable JML-annotated program and associated logs/metrics.}
\footnotesize
\caption{Overview of \name framework} 
\label{fig:overview}
\end{figure*}

Figure~\ref{fig:overview} illustrates the overall architecture of \name, which automates the generation of verifiable JML specifications through a structured, validator-guided workflow. The framework consists of four main components: (1) LLM recommender, (2) primary LLM, (3) iterative refinement, and (4) collaborative LLM, followed by an (5) output stage that produces the final verified specification and execution log. This design enables \name\ to adaptively select appropriate models, refine specifications using validator feedback, and coordinate multiple LLMs to generate correct and verifiable results. The overall procedural flow is summarized in Algorithm~\ref{algo:autorespec}, which formalizes the iterative refinement process used by both LLMs. The notation for the Algorithm is mentioned in Table~\cite{algorithmnotation}.

\begin{algorithm}[t]
\caption{Simplified high-level \name\ workflow}
\label{algo:autorespec}

\KwIn{Java code $\mathcal{J}$, model recommender $\mathcal{R}$, prompt generator $\mathcal{P}$, 
few-shot examples $\mathcal{F}$, system prompts $\mathcal{P}_\text{ps}$ and $\mathcal{P}_\text{cs}$, 
validator $\mathcal{V}$, error analyzer $\mathcal{E}_A$, iteration limits $I_\text{Pmax}$ and $I_\text{Cmax}$.}
\KwOut{Annotated Java code $\mathcal{J_A}$, final validation errors $\mathcal{E}$, iteration statistics, and logs $\mathcal{L}$.}

\SetKwFunction{RefinePhase}{RefinePhase}
\SetKwProg{Fn}{Function}{:}{}
\Fn{\RefinePhase{$\mathcal{J}$, $\mathcal{M}$, $\mathcal{P}$, $\mathcal{P}_s$, $\mathcal{F}$, $I_\text{max}$}}{
    $prompt_{(0)} \gets \mathcal{P}.\texttt{initial}(\mathcal{P}_s, \mathcal{F}, \mathcal{J})$\;
    $\mathcal{L} \gets \texttt{initLog}(prompt_{(0)})$\;
    $prompt \gets prompt_{(0)}$ \;
    \For{$i \gets 1$ \KwTo $I_\text{max}$}{
        $\mathcal{J_A} \gets \mathcal{M}(prompt)$ \;
        $\mathcal{E} \gets \mathcal{V}(\mathcal{J_A})$\;
        $\mathcal{L}.\texttt{append}(\mathcal{J_A}, \mathcal{E})$\;
        \If{$\mathcal{E} = \emptyset$}{
            \KwRet{$\mathcal{J_A}, \emptyset, i, \mathcal{L}$}\;
        }
        $prompt_g \gets \mathcal{P}.\texttt{guidance}(\mathcal{E}_A(\mathcal{E}))$\;
        $prompt \gets \mathcal{P}.\texttt{refine}(prompt, \mathcal{J_A}, \mathcal{E}, prompt_g)$\;
        $\mathcal{L}.\texttt{append}(prompt)$\;
    }
    \KwRet{$\mathcal{J_A}, \mathcal{E}, I_\text{max}, \mathcal{L}$}\;
}

\BlankLine
$T \gets \mathcal{R}.\texttt{getProgramType}(\mathcal{J})$\;
$(\mathcal{M}_P, \mathcal{M}_C) \gets \mathcal{R}.\texttt{getModels}(T)$\;

$(\mathcal{J_A}, \mathcal{E}, I_\text{Pused}, \mathcal{L}_P) 
   \gets$ \RefinePhase{$\mathcal{J}, \mathcal{M}_P, \mathcal{P}, \mathcal{P}_\text{ps}, \mathcal{F}, I_\text{Pmax}$}\;

\If{$\mathcal{E} \neq \emptyset$}{
  $(\mathcal{J_A}, \mathcal{E}, I_\text{Cused}, \mathcal{L}_C) 
   \gets$ \RefinePhase{$\mathcal{J_A}, \mathcal{M}_C, \mathcal{P}, \mathcal{P}_\text{cs}, \mathcal{F}, I_\text{Cmax}$}\;
  $\mathcal{L} \gets \mathcal{L}_P \cup \mathcal{L}_C$\;
}
\Else{
  $\mathcal{L} \gets \mathcal{L}_P$\;
}

\KwRet{$\mathcal{J_A}, \mathcal{E}, I_\text{Pused}, I_\text{Cused}, \mathcal{L}$}\;

\end{algorithm}

Given an input Java program $\mathcal{J}$, the \llmrecommender first analyzes its abstract syntax tree (AST) or equivalent structural representation to infer the program’s type and selects a suitable pair of language models for specification generation. The primary LLM is responsible for the initial synthesis of formal specifications, while the collaborative LLM acts as a fallback when verification fails. The \promptgenerator constructs and manages all prompts used by these models, including initial prompts with few-shot examples and subsequent refinement prompts that incorporate validator feedback. Each LLM iteratively generates, validates, and refines specifications until a verifiable output is produced or the iteration limit is reached. 

When the primary LLM cannot generate a valid specification, \name triggers the collaborative phase. The collaborative LLM receives only the final invalid specification and its associated validation feedback from the primary phase, allowing focused correction without redundant context. Both phases follow the same refinement protocol, ensuring consistency while improving efficiency and verification success. 

By combining adaptive model selection with validator-guided refinement, \name\ achieves an efficient two-phase generation process that progressively converges to verified specifications.

\subsection{Dynamic LLM Selection via Recommendation}
\label{subsec:llm-recommender}

The \llmrecommender serves as an adaptive selection mechanism that identifies the most suitable pair of LLMs for a given program. To determine the appropriate LLMs, the framework first constructs an abstract syntax tree (AST) of the input program to capture its structural hierarchy. Using a rule-based parser, the program type extractor classifies the code into one of five categories~\cite{specgen, xie2017automatic}: \textit{Sequential}, \textit{Branched}, \textit{Single-path Loop}, \textit{Multi-path Loop}, or \textit{Nested Loop}. These categories reflect different control-flow complexities that influence the difficulty of generating the specification.

To empirically establish model preferences for each program type, we conducted a lightweight calibration study prior to full-scale evaluation. For every program category, three representative samples were randomly selected and evaluated across all candidate LLMs using identical prompting and validation configurations. For each model, we computed effectiveness metrics: Success Rate Percentage ($SR$), completeness ($\mathcal{C}$), and efficiency metrics: number of validator calls ($N_{\text{val}}$) and mean generation time. These statistics were then aggregated to form a ranked list of model performance per program type.

Based on this ranking, the \llmrecommender adopts a cost-aware pairing strategy by default: it assigns a smaller, faster model as the primary LLM for initial specification synthesis, and a higher-capacity or costlier model as the collaborative LLM for recovery when verification fails. This configuration balances efficiency and reliability while minimizing redundant computation. The \llmrecommender also determines an appropriate few-shot configuration specifying the number and selection strategy of examples tailored to the LLM. While \name\ uses cost-effectiveness as its default selection policy, users can customize the ranking criteria to prioritize alternative objectives such as accuracy, latency, or energy efficiency, depending on their application requirements.

This recommendation mechanism ensures that each test case is handled by an optimized pair of LLMs, balancing efficiency, model diversity, and success probability rate.

\subsection{Prompt Construction and Refinement}
\label{subsec:prompt-generator}
After the LLM pair and few-shot configuration are determined by the \llmrecommender, the \promptgenerator constructs and refines the input prompts for both models. The \promptgenerator generates the input prompts used by both the primary and collaborative LLMs. For the initial prompt to the primary LLM, \name combines the primary LLM's system message, a configurable number of few-shot examples (as determined by the \llmrecommender described in section~\ref{subsec:llm-recommender}), and a specification generation prompt that includes the input Java code. The few-shot examples are randomly chosen from a curated set of expert-validated specifications. These examples provide structural and formatting cues, enabling the LLM to produce well-structured responses with correctly delimited code blocks that can be programmatically extracted~\cite{brown2020language}. The system prompt, format of the few-shot examples, and the final specification generation prompts are shown in ~\cite{AutoReSpecPrompts}.

The collaborative LLM is invoked only when the primary LLM fails to produce a valid specification. Its objective is to correct the validation error identified by the validator and generate a revised, valid specification. The initial prompts for the collaborative LLM closely resemble those of the primary LLM, with additional instructions. The system prompt for the collaborative LLM includes instructions to synthesize the validation error and produce a corrected specification within a single code block. The initial prompts also include a configurable number of few-shot examples similar to those of the primary LLM. Additionally, it incorporates an initial error feedback prompt instead of a specification generation prompt, as illustrated in ~\cite{AutoReSpecPrompts}. The final failed output from the primary LLM, along with the associated validator error, provides the collaborative LLM with complete context to attempt recovery.

For iterative refinement as described in~\ref{subsec:collaborative-specificaiton-generation}, the \promptgenerator extracts the validation error message generated by the validator (OpenJML) and identifies the corresponding error types. Based on these error types, it generates an error-specific guidance message with refinement examples or falls back to a generic guidance message. This guidance message is then combined with the previously generated invalid JML-annotated code and the validation error to form the next iterative prompt.

Additionally, the constructor retains previous prompt messages to maintain conversational continuity across iterations. LLMs tend to hallucinate more, incur higher computational costs, and exhibit slower response times as the number of input tokens increases. This results from the quadratic attention complexity in transformer architectures, making long-context inference significantly more expensive~\cite{10.1145/3466752.3480125, ji-etal-2023-towards}. To reduce the context window for proprietary models, the \promptgenerator also employs automated prompt truncation to remain within the model constraints. \name retains only the last few prompts or truncates prompts to 4000 tokens, whichever comes first, ensuring both compactness and contextual relevance during iterative refinement.

\subsection{Collaborative LLMs for Iterative Specification Generation}
\label{subsec:collaborative-specificaiton-generation}

After the prompt templates are defined by the \promptgenerator, \name executes the specification generation process in two sequential phases, led by a primary and a collaborative LLM. Hallucinations remain a significant challenge in LLMs, but they can be mitigated through conversational prompting, which helps align model outputs with user intent and maintain contextual consistency~\cite{zhang2023siren}. Following the prompting strategy of Xia et al.~\cite{xia2024automated}, \name interacts with LLMs in a conversational manner to generate formal specifications. By iteratively incorporating validation errors and relevant guidance messages into the prompts, the models refine their outputs over successive turns. This approach not only helps reduce syntax errors but also provides a mechanism for addressing semantic verification errors, often improving specification accuracy and completeness.

The specification generation process begins with the primary LLM recommended by the \llmrecommender for the queried program type. The initial prompt for this LLM is generated by \promptgenerator and sent to the selected model to produce a specification. Each generated specification is immediately verified; if validation fails, the \promptgenerator parses the error messages, retrieves relevant guidance, and reconstructs the next prompt for iterative refinement. This conversational loop continues until a valid specification is produced or the primary iteration limit is reached.

If the primary LLM fails to generate a correct specification within its allotted iterations, \name performs a controlled memory reset before invoking the collaborative LLM. This resets the full conversational history from the primary phase while retaining only the latest invalid specification and its corresponding validation errors. The \promptgenerator then constructs an initial collaborative prompt combining this minimal context with few-shot examples and the associated verification feedback. The collaborative LLM, also selected by the \llmrecommender, begins reasoning afresh to refine the failed specification. Since it receives only task-relevant input rather than the full conversation history, the collaborative model avoids bias or context drift while maintaining efficiency. If the collaborative LLM succeeds, the verified specification is returned; otherwise, the final invalid output and verification errors are provided as a starting point for manual refinement.
\section{Experimental Design}
\label{sec:experimantaldesign}
We evaluate \name by formulating the following research questions (RQs) and outlining our approach to answering them:

\begin{itemize}
    \item \textbf{RQ1: How applicable is collaborative-LLM strategy for generating specifications?} 
    \item \textbf{RQ2: How effective is \name in generating formal specifications compared to state-of-the-art LLM-based specification generation tools?} 
    \item \textbf{RQ3: How efficient is \name in generating specifications compared to existing automated tools?}
    \item \textbf{RQ4: How effectively does \name resolve verification failures compared to state-of-the-art approaches?} 
\end{itemize}

\subsection{Implementation}
\label{subsec:implementation}

We used Ollama to run open‐source LLMs locally and accessed proprietary models (Claude 3.7 Sonnet, GPT-4o) via Anthropic and OpenAI APIs. Based on prior findings~\cite{brown2020language}, we set the temperature to 0.4 for all models. \name currently targets Java, prompting LLMs to produce JML annotations that OpenJML 21.0.8 verifies using z3 4.3.1 as its SMT solver~\cite{openjml2108}. Because OpenJML can hang or fail, we imposed a 180-second timeout per verification. For completeness measurement, we used Major 3.0.1~\cite{just2014major} and applied EMS~\cite{kushigian2024equivalent} to filter out equivalent mutants, ensuring only semantically distinct faults are counted~\cite{le2025can}. All experiments were run on a 28-core Intel® Xeon® W-3465X CPU (2.50 GHz), 256 GB of RAM, and an NVIDIA RTX 6000 Ada Generation GPU under Ubuntu 24.04.2 LTS. Except where noted (Major requiring OpenJDK 11), we used OpenJDK 21.0.  

\subsection{Dataset}
\label{subsec:dataset}
Following prior work~\cite{alshnakat2020constraint}, to thoroughly evaluate \name, we develop a new benchmark dataset, \leaderboardBench, consisting of challenging Java classes drawn from SpecGenBench, SV-COMP~\cite{specgen,svcomp}, and real-world cases from GitHub issues. Importantly, many benchmark instances are multi-method classes rather than only single-function programs, which helps assess whether \name scales beyond single-procedure settings. From SpecGenBench, we select 26 programs that LLMs struggle with most in our preliminary experiments. We additionally extract 29 SV-COMP programs featuring loop-intensive control flow. To further evaluate real-world applicability and reduce the likelihood of training-set overlap, we include 17 new multi-method Java classes curated from OpenJML GitHub issues~\cite{openjmlrepo}, reflecting complex verification scenarios encountered in practice. Overall, \leaderboardBench spans diverse control-flow patterns and common data types (e.g., arrays and strings), and includes varied specification constructs (e.g., postconditions and loop invariants) with both linear and nonlinear relationships among variables.

\subsection{Evaluation Metrics}
\label{subsec:evaluation metrics}
Following previous works~\cite{specgen, le2025can}, we utilize the metrics of Number of Passes, Success Rate, Success Probability, Number of Verifier Calls, and Completeness to assess \name.

\textbf{Number of Passes ($NP$):}
The number of passes ($NP$) counts the number of programs in the dataset for which at least one generated specification is successfully verified within the allotted budget across trials~\cite{specgen} (i.e., the verifier reports no errors). We define the \textit{Success Rate} ($SR$) as the dataset-normalized form of this metric: $SR = \frac{NP}{|\mathcal{D}|}$, where $|\mathcal{D}|$ is the dataset size.

\textbf{Success Probability ($SP$):} 
The Success Probability metric captures how often an LLM‐based approach produces a verifiable specification when run multiple times on the same program. Because LLM outputs can vary from run to run~\cite{ouyang2022training}, we execute each program ten times and record the fraction of runs that pass verification. This average success probability summarizes a model’s reliability across the dataset.  

\textbf{Number of Verifier Calls ($N_{\text{val}}$):} 
The Number of Verifier Calls metric tracks the number of times the verifier is invoked during specification generation. It serves as a proxy for computational effort: fewer calls indicate that a technique needs fewer refinement attempts to produce a correct specification. For \name, we average the sum of primary and collaborative verifier invocations over multiple trials and programs, yielding a single value that reflects overall verifier usage and efficiency.

\textbf{Completeness ($\mathcal{C}$):} 
The Completeness metric~\cite{formalbench} evaluates how well a specification detects injected faults via mutation testing. For each verified specification, we generate a set of non‐equivalent mutants and count how many trigger verifier errors. A higher completeness score indicates broader fault coverage. Averaging over multiple trials and programs yields an overall completeness percentage.

\section{Results and Evaluation}
\label{sec:results-evaluation}

\subsection{RQ1: Applicability of the Collaborative LLM Strategy}
\label{RQ1}
\textit{Experimental Setup:} We evaluate both individual and collaborative LLM configurations for formal specification generation on SpecGenBench~\cite{specgen}. We report the number of passes ($NP$), which indicates how many tasks successfully passed validation from the dataset, as well as Success Probability ($SP$), and completeness ($\mathcal{C}$), both measured over validated programs only. All individual models are tested with zero-, two-, and four-shot prompting, while the top open-source models (Llama 3 (8B), Phi 4 (14B), and Gemma 3 (27B)) are also assessed with conversational prompting. Overall, we tested 120 issues with ten trials each to compute $SP$, and $\mathcal{C}$. To ensure a fair comparison between conversational and collaborative settings, we set the maximum number of validator calls to 10. For collaborative prompting, the budget is evenly divided between the primary and collaborative LLMs. In the Dynamic LLM$\dagger$ configuration, we use ten iterations for both the primary and collaborative LLMs. For dynamic LLM selection, we select the highest-scoring model configuration for each program type. For Sequential and Branched programs, we select Gemma~3, with GPT-4o (Sequential) and Claude-3.7-Sonnet (Branched) as counterparts; for Single- and Multi-path loops, we select Llama~3 and GPT-4o; and for Nested loops, we select Llama~3 and Claude-3.7-Sonnet. To keep our study's costs under budget, following prior work~\cite{10.1145/3643763}, we do not evaluate closed-source LLMs in the individual conversational setting and instead use the individual evaluation setting as a basis for dynamically selecting the LLM. 

\renewcommand{\arraystretch}{1.15}
\begin{table}[!htbp]
  \centering
  \caption{Performance of LLM-generated specifications on the SpecGenBench dataset under different prompting settings in \name}
  \begin{adjustbox}{width=\columnwidth}
  \begin{tabular}{@{}p{0.2cm} l l c c c@{}}
    \toprule
    \textbf{} & \textbf{Model} & \textbf{Prompting} & \textbf{$NP$} & \textbf{$SP$ (\%)} & \textbf{$\mathcal{C}$ (\%)} \\
    \midrule

    \multirow{24}{*}{\rotatebox{90}{\textbf{Individual}}}
      & \multirow{3}{*}{Claude-3.7-Sonnet} 
        & Zero-shot & 62 & 69.84 & 89.84 \\
      & & Two-shot & 74 & 74.46 & 91.02 \\
      & & Four-shot & 76 & 74.74 & 89.86 \\
      \cmidrule(lr){2-6}

      & \multirow{3}{*}{GPT-4o}
        & Zero-shot & 74 & 75.27 & 85.62 \\
      & & Two-shot & 76 & 78.82 & 81.79 \\
      & & Four-shot & 79 & 76.58 & 84.05 \\
      \cmidrule(lr){2-6}

      & \multirow{4}{*}{Llama 3 (8B)}
        & Zero-shot & 67 & 49.55 & 10.74 \\
      & & Two-shot & 53 & 37.36 & 47.72 \\
      & & Four-shot & 57 & 40.88 & 42.30 \\
      & & Conv. & 103 & 64.66 & 29.92 \\
      \cmidrule(lr){2-6}

      & \multirow{4}{*}{Gemma 3 (27B)}
        & Zero-shot & 41 & 63.90 & 36.51 \\
      & & Two-shot & 62 & 58.87 & 79.54 \\
      & & Four-shot & 61 & 59.67 & 82.01 \\
      & & Conv. & 77 & 69.61 & 77.94 \\
      \cmidrule(lr){2-6}

      & \multirow{4}{*}{Phi 4 (14B)}
        & Zero-shot & 45 & 42.22 & 34.81 \\
      & & Two-shot & 58 & 54.14 & 75.80 \\
      & & Four-shot & 57 & 54.91 & 72.97 \\
      & & Conv. & 85 & 54.82 & 63.17 \\
      \cmidrule(lr){2-6}

      & \multirow{3}{*}{Llama 3.3 (70B)}
        & Zero-shot & 43 & 37.91 & 58.49 \\
      & & Two-shot & 60 & 44.50 & 87.44 \\
      & & Four-shot & 54 & 55.93 & 89.58 \\
      \cmidrule(lr){2-6}

      & \multirow{3}{*}{Mistral (7B)}
        & Zero-shot & 12 & 10.00 & 6.38 \\
      & & Two-shot & 40 & 29.25 & 56.31 \\
      & & Four-shot & 44 & 26.59 & 60.21 \\
    \midrule

    \multirow{3}{*}{\rotatebox{90}{\textbf{Collab.}}}
      & Llama 3 \& Llama 3  
      & Four-shot & 105 & 23.10 & 21.38 \\
      & GPT-4o \& GPT-4o 
      & Four-shot & 94 & 62.42 & \textbf{84.84} \\
      & \textbf{Dynamic LLM$\dagger$}
      & \textbf{Few-shot} & \textbf{119} & \textbf{69.32} & 60.33 \\
    \bottomrule
  \end{tabular}
  \end{adjustbox}
  \label{tab:RQ1}
  \begin{flushleft}
    \footnotesize
    \textbf{Note:} $\dagger$~Primary \& collaborative iterations~=~10; Collab.~=~Collaborative Prompting; Conv.~=~Conversational; $NP$~=~Passes; $SP$~=~Success Probability; $\mathcal{C}$~=~Completeness.
    \vspace{-2.5em}
  \end{flushleft}
\end{table}

\textit{Result Analysis:} From Table~\ref{tab:RQ1}, collaborative prompting achieves higher overall verification coverage and produces more consistent specifications compared to other configurations. \name with collaborative prompting and dynamic model selection reaches an $NP$ of 119 out of 120 programs, surpassing the state of the art, SpecGen~\cite{specgen}, with 100. 

A controlled ablation on Llama 3 (8B) evaluates zero, two, four-shot, conversational, and collaborative prompting. Zero-shot performance is poor ($NP$ = 67, $SP$ = 49.55\%, $\mathcal{C}$ = 10.74\%) because the model fails to infer structure without exemplars. Two and four-shot prompting improves both $SP$ and $\mathcal{C}$ (up to 57 passes and 42.30\%). Individual conversational prompting further increases the number of passes ($NP$ = 103, $SP$ = 64.66\%), while collaborative prompting achieves the highest $NP$ (105) with moderate completeness ($\mathcal{C}$ = 21.38\%). This confirms that collaboration, even between identical models, improves the number of passes without additional iterations. The dynamic LLM$\dagger$ configuration extends this by adaptively selecting between two and four-shot prompts, achieving the best $NP$ (119) and a competitive $SP$ (69.32\%), indicating improved consistency and robustness across diverse program types.

For proprietary models, GPT-4o and Claude 3.7 Sonnet perform strongly under individual prompting. GPT-4o attains $NP$ = 79, $SP$ = 76.58\%, and $\mathcal{C}$ = 84.05\%, while Claude 3.7 achieves $NP$ = 74, $SP$ = 74.46\%, and the highest $\mathcal{C}$ = 91.02\%. In the collaborative–conversational setup, GPT-4o improves to $NP$ = 94 and $\mathcal{C}$ = 84.84\%. 

Across open-source models, we observe consistent gains from zero to few-shot and conversational prompting. Llama 3 (8B), Gemma 3 (27B), and Phi 4 (14B) benefit from structured prompting and validator-guided refinement. Llama 3.3 (70B) achieves high completeness under four-shot prompting but lower $SP$ (55.93\%) and $NP$, suggesting that model size alone does not guarantee stronger generalization. Mistral (7B) performs weakest across all metrics, highlighting the importance of scale and prompt design.

Overall, collaborative prompting with dynamic model selection achieves the best balance across $NP$, $SP$, and $\mathcal{C}$, showing that structured collaboration and adaptivity improve verification outcomes without increasing iteration count.

\finding[Answer to RQ1]{The collaborative-LLM strategy is effective for specification generation. \name achieves 119 verified passes compared to SpecGen’s 100, with a 69.32\% success probability and 60.33\% completeness. These results indicate that structured collaboration and adaptive model selection improve verification outcomes and practicality without increasing the iteration budget.}

\subsection{RQ2: Effectiveness of \name}
\label{RQ2}

\textit{Experimental Setup:} 
To assess the effectiveness of \name, we evaluate it against two state-of-the-art LLM-based specification generation techniques, SpecGen and FormalBench. Experiments are conducted on \leaderboardBench using ${SR}{\%}$, $SP$, and $\mathcal{C}$, where $SP$ and $\mathcal{C}$ are computed only for programs with valid JML annotations. For statistical validation, we use McNemar’s test for paired pass/fail outcomes~\cite{mcnemar1947note}, and the paired Wilcoxon signed-rank test for per-program $SP$ and $\mathcal{C}$~\cite{wilcoxon1945individual}. To ensure a fair runtime comparison, we cap SpecGen’s validator calls at 20 and reduce the inter-call delay from 20s to 1s; \name uses the dynamic model selection configuration.

\begin{figure}[!htbp]
\centering
\includegraphics[width=\linewidth,trim={0cm 0cm 0cm 0cm},clip]{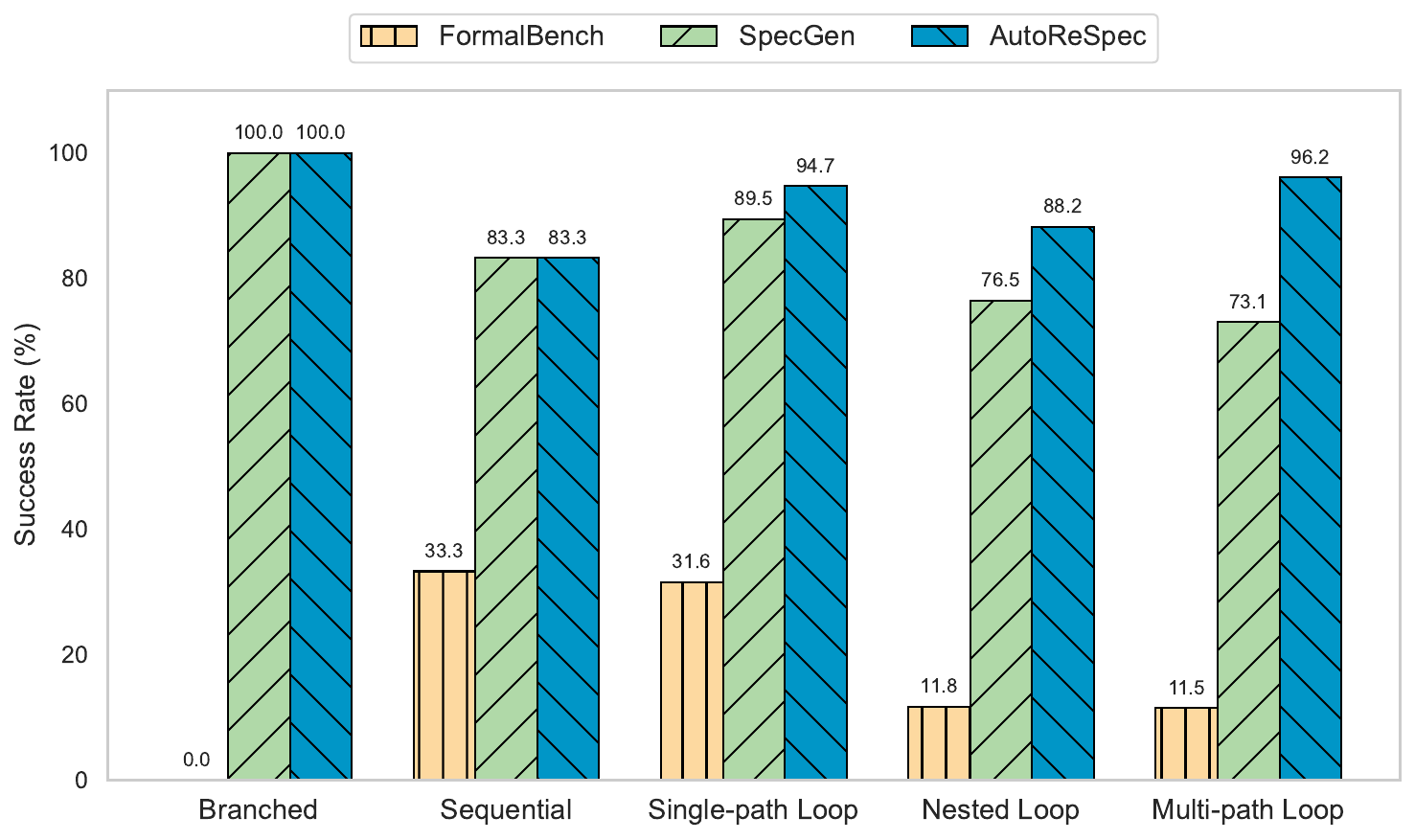}
\Description{Grouped bar chart of pass percentage (SR\%) by program type (e.g., Sequential, Branched, Single-path Loop, Multi-path Loop, Nested Loop) comparing \name, SpecGen, and FormalBench. Across all categories, \name attains the highest or tied-highest pass rate, with the largest margins on loop-heavy program types.}
\footnotesize
\caption{Pass percentage (${SR}{\%}$) across program types for state-of-the-art specification generation techniques.}
\label{fig:RQ2-pass-percentage}
\end{figure}

\begin{figure}[!htbp]
\centering
\includegraphics[width=\linewidth,trim={0cm 0cm 0cm 0cm},clip]{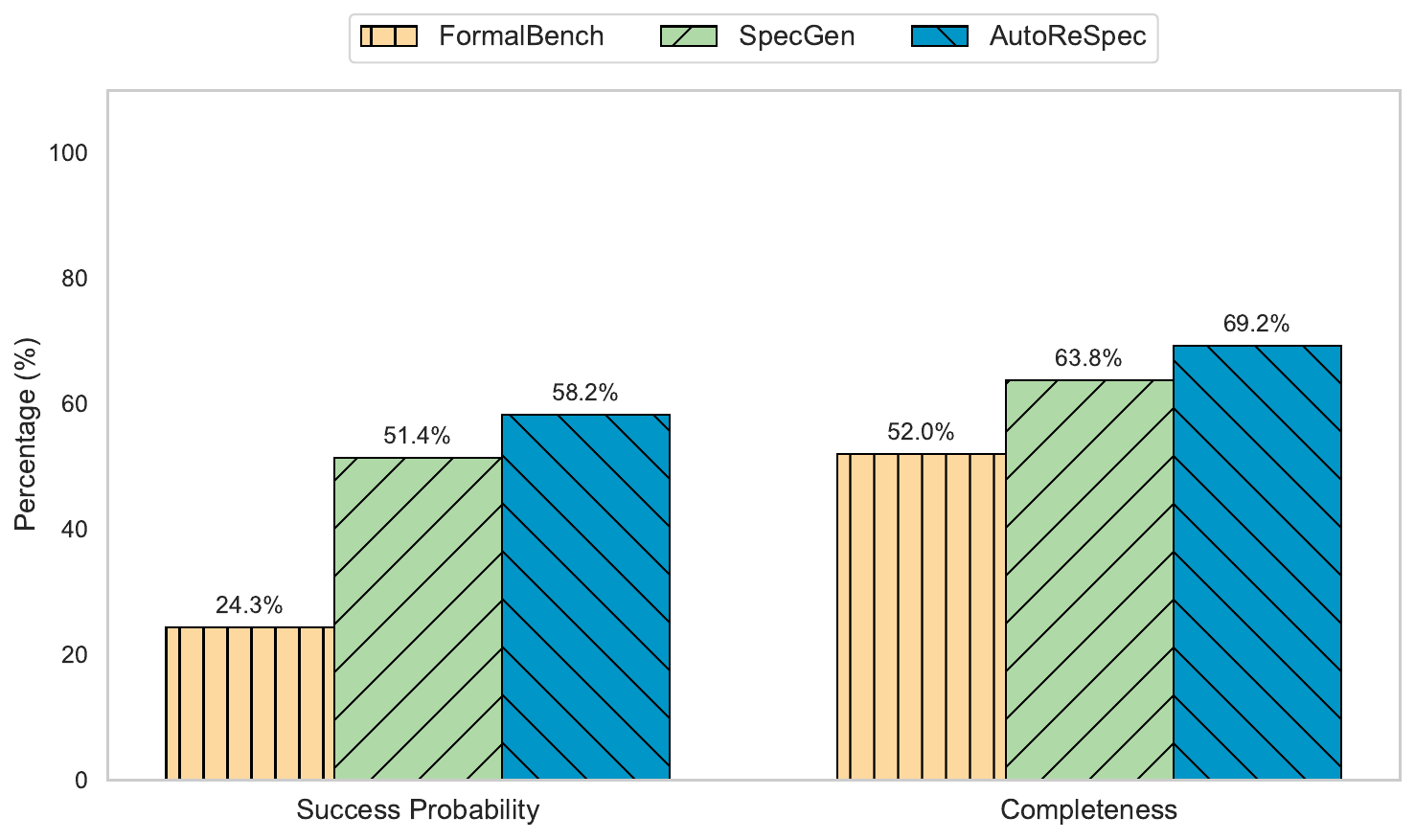}
\Description{Bar chart comparing three techniques (\name, SpecGen, and FormalBench) on two averaged metrics: success probability (SP) and completeness (C). \name shows the highest average success probability and completeness, SpecGen is second on both metrics, and FormalBench has the lowest values.}
\footnotesize
\caption{Average Success Probability and Completeness across LLM-based specification generation techniques.} 
\label{fig:RQ2-sp-completeness}
\end{figure}

\textit{Result Analysis:} 
Figure~\ref{fig:RQ2-pass-percentage} presents the ${SR}{\%}$ of all techniques on \leaderboardBench, showing that \name matches or exceeds both SpecGen and FormalBench across all program categories. 
\name achieves notable gains in program types that are generally challenging for LLMs, particularly those involving loops. 
For Multi-path Loops, \name produced 25 valid specifications out of 26, compared to SpecGen’s 19 and FormalBench’s 3 (a 23.1\% improvement over the best baseline). 
For Nested Loops, it achieved 15 passes out of 17, exceeding SpecGen (13) and FormalBench (2) by 11.8\%. 
In Single-path Loops, \name verified 18 of 19 programs, surpassing SpecGen (17) and FormalBench (6) with a 5.3\% gain. 
On less complex categories, \name matched SpecGen’s performance, successfully passing all 4 Branched programs and 5 of 6 Sequential programs. 
These results indicate that \name performs robustly across diverse program structures, with consistent improvements on programs featuring complex control flow.

Figure~\ref{fig:RQ2-sp-completeness} reports the average success probability ($SP$) and completeness ($\mathcal{C}$) across techniques. Leveraging dynamic model selection and collaborative prompting, \name attains the highest average $SP$ (58.2\%), exceeding SpecGen (51.4\%) and FormalBench (24.3\%). \name also achieves the highest average completeness, which is approximately 5.4\% higher than SpecGen and substantially higher than FormalBench. Overall, \name yields more reliable and comprehensive specifications in aggregate, with statistically significant gains over FormalBench.

Since all techniques are evaluated on the same programs, we apply paired statistical tests. For pass/fail coverage (SR/NP), McNemar’s test shows \name improves over SpecGen ($p{=}0.049$; 13 vs.\ 4 discordant passes; $\Delta SR{=}0.125$, 95\% CI [0.014, 0.236]) and strongly outperforms FormalBench ($p{<}10^{-15}$; 54 vs.\ 0; $\Delta SR{=}0.75$, 95\% CI [0.653, 0.847]). For $SP$ and $\mathcal{C}$, paired Wilcoxon tests show significant gains over FormalBench ($p{<}10^{-11}$), but not over SpecGen ($p{=}0.067$ for $SP$, $p{=}0.260$ for $\mathcal{C}$).

\finding[Answer to RQ2]{\name generates valid specifications for 67 of 72 programs, compared to SpecGen’s 58 and FormalBench’s 13. It achieves the highest overall average success probability and completeness, with statistically significant gains over FormalBench and significantly higher pass coverage than SpecGen.}

\subsection{RQ3: Efficiency of \name}
\label{RQ3}

\textit{Experimental Setup:} 
We evaluate the efficiency of \name against SpecGen and FormalBench using the same benchmark programs and experimental setup described in Section~\ref{RQ2} to ensure consistency.  We report the average evaluation time ($T_{\text{eval}}$) and number of validator calls ($N_{\text{val}}$). To maintain fairness, SpecGen’s delay between successive LLM calls is reduced from 20 seconds to 1 second, and its validator calls are capped at 20. For \name, we employ the \textit{dynamic model selection configuration}. In this evaluation, since all open-source LLMs were executed on our local hardware without per-request API charges, we do not report monetary cost for open-source models.

\begin{figure}[!htbp]
\vspace{-1.5em}
\centering
\includegraphics[width=\linewidth,trim={0cm 0cm 0cm 0cm},clip]{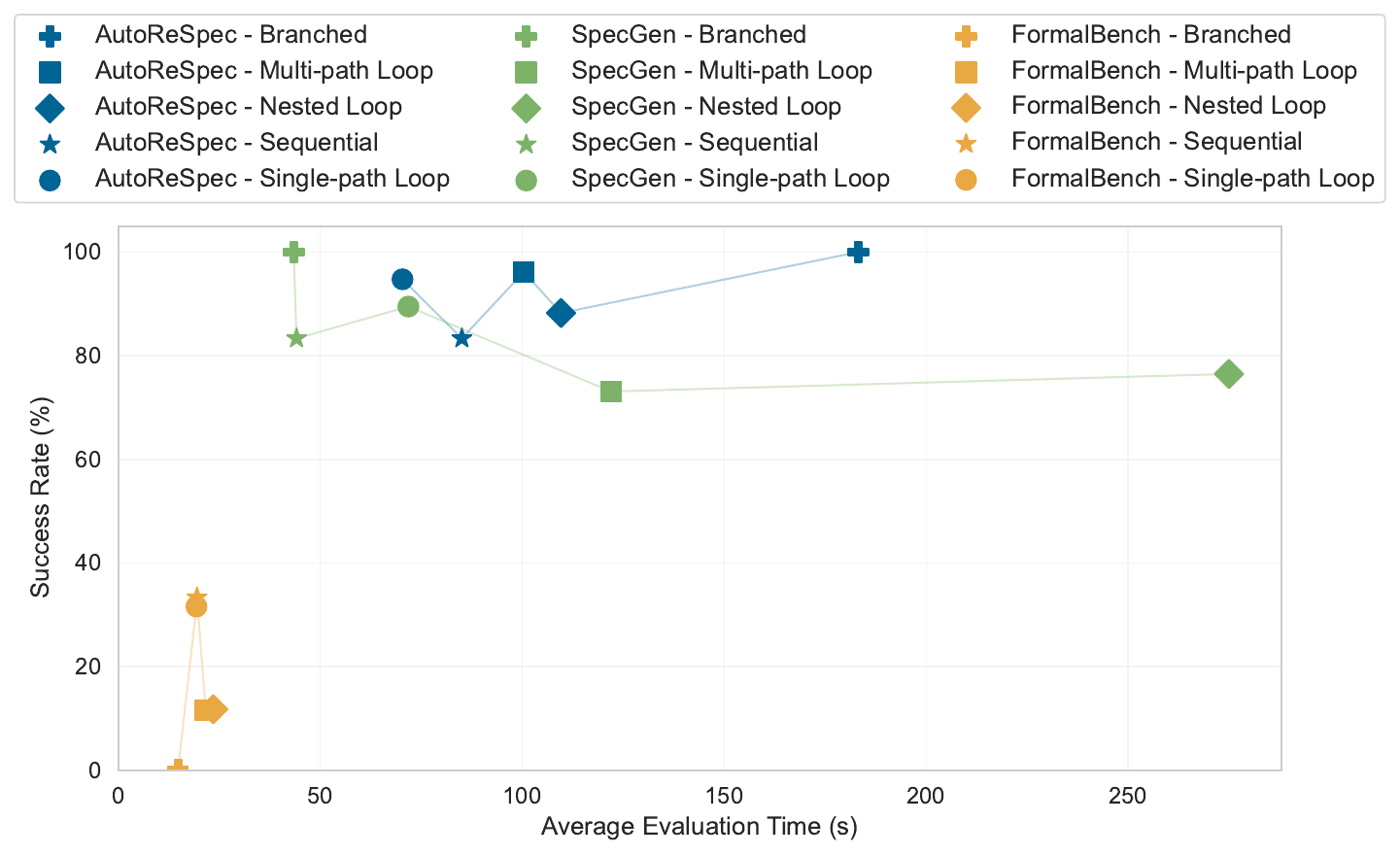}
\Description{Visualization of average evaluation time and success rate percentage across different program types for LLM-based specification generation techniques.}
\footnotesize
\caption{Average evaluation time and success rate percentage across different program types of LLM-based specification generation techniques.}
\vspace{-2em}
\label{fig:RQ4-time-vs-pp}
\end{figure}

\textit{Result Analysis:} 
As shown in Figure~\ref{fig:RQ4-time-vs-pp}, \name achieves a balanced trade-off between runtime efficiency and verification success among the three techniques. Although FormalBench has the shortest average runtime (19.8 s), its low success probability (18.1\%) limits practical use. Its validator calls are capped at 3, and it uses 2.38 on average in our experiments. SpecGen requires 111.3 s on average and achieves 80.1\% success probability, while \name completes evaluation in 109.7 s with a higher success probability of 94.5\%. Despite using slightly more validator calls (12.4 vs. 8.2), the adaptive selection mechanism in \name reduces redundant refinements, resulting in a marginally faster overall runtime. Although FormalBench has the shortest average runtime (19.8 s), its low success probability (18.1\%) limits practical use. FormalBench also operates under a small validator-call budget (capped at 3), and in our experiments, it used 2.38 validator calls on average.

In terms of API costs for proprietary models, \name averages \$0.13 per Java class, which is lower than SpecGen (\$0.16) and FormalBench (\$0.17). Costs scale moderately with program complexity, ranging from \$0.11 for sequential programs to \$0.25 for multi-path loops, and are capped at \$0.31 for the most complex cases. Because \name prioritizes open-source models and invokes proprietary ones only when necessary, it maintains verification quality while controlling overall cost. 

These results suggest that efficiency gains primarily arise from the dynamic LLM selection mechanism, which reduces redundant validation loops by choosing suitable prompt configurations for each task.  Runtime grows approximately linearly with validator calls and remains bounded by the iteration cap, indicating good scalability to larger codebases.

\finding[Answer to RQ3]{\name achieves runtime efficiency comparable to SpecGen (109.7 s vs. 111.3 s) while using slightly more validator calls on average (12.4 vs. 8.2). Despite this, it remains marginally faster overall and incurs a lower average API cost (\$0.13 per Java class), indicating practical efficiency, cost-effectiveness, and scalability.}

\subsection{RQ4: Error Resolution Capabilities of \name}
\label{RQ4}

\textit{Experimental Setup:} 
To better understand the challenges each technique faces during specification generation, we conducted a verification error type–level struggle analysis. Each verification error type (e.g., \textit{Postcondition}, \textit{LoopInvariantBeforeLoop}) represents a distinct class of JML validation failure. We compute the struggle ratio as the frequency of each error type relative to the total number of validator calls across all specifications, quantifying how persistently difficult each category is for a given technique. For \name, we used the \textit{dynamic model selection configuration}. 
We also manually examined cases where \name failed to produce verifiable specifications to identify recurring error patterns and root causes.

\begin{figure}[!htbp]
\centering
\includegraphics[width=\linewidth,trim={0cm 0cm 0cm 0cm},clip]{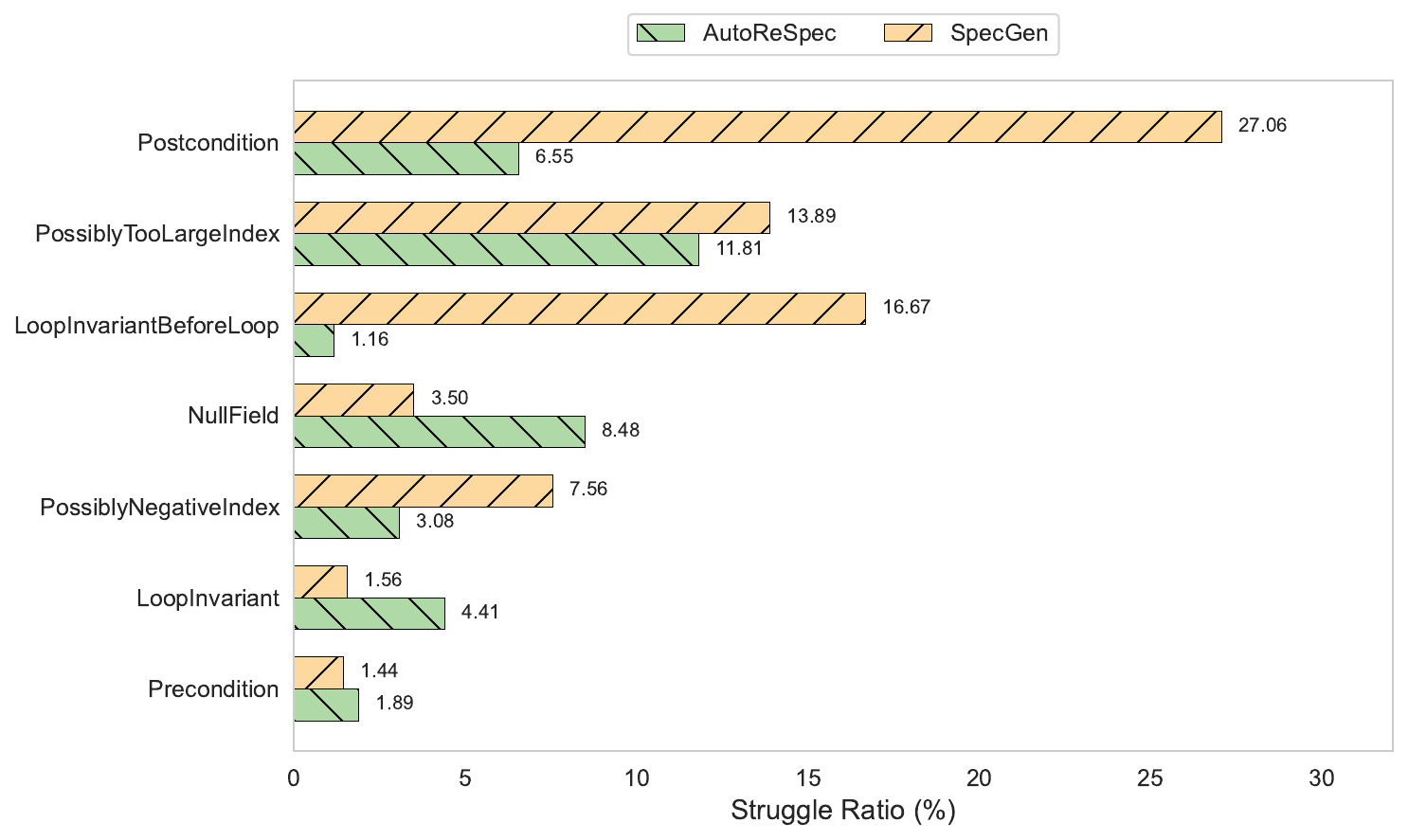}
\Description{Horizontal bar chart showing the most common OpenJML verification error categories and their struggle ratios (percentage of programs where generation fails due to that error), comparing \name and SpecGen. For each error type, two bars indicate the struggle ratio for each technique, with \name generally exhibiting lower struggle ratios than SpecGen across the top error categories.}
\footnotesize
\caption{Top verification error types and struggle ratios (\%) for \name and SpecGen.} 
\label{fig:RQ4-struggle}
\end{figure}

\textit{Result Analysis:} 
Figure~\ref{fig:RQ4-struggle} compares struggle ratios for seven common verification error types across SpecGen and \name. Postcondition errors remain the most persistent overall: SpecGen records a 27.06\% struggle ratio, while \name reduces this to 6.55\%, indicating stronger handling of functional correctness constraints. Similarly, for \textit{LoopInvariantBeforeLoop}, \name achieves 1.16\% compared to SpecGen’s 16.67\%, suggesting more consistent reasoning over iterative control structures. For index-related categories (\textit{PossiblyTooLargeIndex} and \textit{PossiblyNegativeIndex}), \name achieves lower ratios (11.81\% vs.\ 13.89\% and 3.08\% vs.\ 7.56\%), reflecting improved handling of array bounds. However, \name shows a higher ratio for \textit{NullField} errors (8.48\% vs.\ 3.50\%), indicating an area where null-safety inference could be enhanced. For other types, such as \textit{LoopInvariant} and \textit{Precondition}, both techniques exhibit similarly low ratios.

A closer inspection of the few failed cases revealed recurring issues mainly related to JML syntax and complex control dependencies. The most frequent failure patterns include \textit{Precondition}, \textit{Assert}, \textit{LoopInvariantBeforeLoop}, \textit{Postcondition}, and \textit{InvariantLeaveCaller}. In one representative case (\textit{TransposeMatrix}), a catastrophic JML internal failure occurred due to double rewriting of identifier references. These failures often occur in programs with deeply nested loops and interdependent postconditions, where current LLMs still struggle to maintain consistency between inferred invariants and postconditions.

Overall, the results suggest that \name mitigates several persistent verification challenges, especially those involving postconditions and loop invariants, while leaving room for improvement in null-safety reasoning.
\finding[Answer to RQ4]{\name achieves lower struggle ratios across most verification error types compared to SpecGen, particularly for Postcondition (27.06\% vs.\ 6.55\%) and LoopInvariantBeforeLoop (16.67\% vs.\ 1.16\%). These results indicate improved error resolution capacity through LLM-guided refinements while maintaining robustness across complex program structures.}

\textbf{Limitation}
Because \name relies on LLMs’ code understanding, it inherits their weaknesses. LLMs can misinterpret complex control flow~\cite{le2025can}, sometimes return altered or incomplete code (necessitating expert review), and may omit code entirely, which adds post-processing overhead. On the verification side, OpenJML can produce catastrophic errors on programs with matrix operations, and its SMT solver (Z3) may time out on complex proof obligations, yielding inconclusive results. Finally, our completeness measurement uses Major with Equivalent Mutant Suppression (EMS)~\cite{kushigian2024equivalent}, but EMS does not catch every equivalent mutant, slightly lowering completeness scores and increasing evaluation time.
\section{Related Work}
\label{sec:related-works}
Recent advances in LLMs have revived interest in automated specification generation and verification. This section scopes our work within three areas of related research: (1) LLM-assisted specification inference, (2) traditional static and dynamic approaches to specification mining, and (3) evaluation frameworks and benchmarks for verifying specification quality.

\subsection{Specification Generation with LLMs}
LLMs have shown promise in software engineering tasks such as code generation~\cite{zeng2022extensive}, summarization~\cite{ahmed2022few}, and defect prediction~\cite{hou2024large}, motivating their application to formal specification inference~\cite{pei2023can,chakraborty2023ranking}. SpecGen~\cite{specgen} employs conversational prompting with few-shot examples and a mutation-based process that applies heuristic selection to obtain verifiable JML specifications. Other recent studies~\cite{wen2024enchanting,pei2023can,chakraborty2023ranking} explore integrating LLMs with static analysis or ranking mechanisms to improve invariant quality, but they remain limited to single-model or fixed-prompt configurations. \name advances this direction through collaborative prompting for cross-model refinement and adaptive model selection based on program type and validator feedback, demonstrating that prompt design, fallback ordering, and iterative refinement critically influence verification success and computational efficiency.

\subsection{Traditional Specification Inference}
Prior to LLMs, specification inference relied on dynamic and static analysis. Tools like Daikon~\cite{ernst2007daikon} mine invariants from execution traces, while constraint solvers and abstract interpretation methods target lightweight formal properties~\cite{alshnakat2020constraint, cousot2013automatic}. Some hybrid approaches incorporate symbolic execution or deductive inference~\cite{chakraborty2023ranking, molina2021evospex}, though these often operate under constrained grammars or language subsets (e.g., C/Frama-C). Others, like EvoSpex~\cite{molina2021evospex}, introduce ranking or mutation strategies but lack feedback loops from verifiers or adaptive prompt refinement. \name departs from these patterns by integrating LLMs with a verifier-guided conversational interface. We introduce a dynamic prompt constructor that adapts based on failure type and validator output, addressing a limitation of prior static and batch-mode approaches. This makes \name uniquely suited for open-ended code with real-world variability.

\subsection{Evaluation Frameworks and Benchmarks}
SpecGen~\cite{specgen} and FormalBench~\cite{formalbench} offer valuable foundations for evaluating LLM-generated specifications. SpecGen primarily measures verification accuracy under few-shot prompting, while FormalBench's toolkit broadens the scope to reasoning metrics such as consistency, completeness, and robustness across diverse prompting strategies. However, these frameworks focus on single-model settings and fixed prompts, providing limited insight into how models recover from verification failures, adapt across iterations, or balance accuracy and efficiency. Recent agent-based evaluation efforts, such as CodeVisionary~\cite{wang2025codevisionaryagentbasedframeworkevaluating} and RepoMasterEval~\cite{wu2024repomastereval}, further emphasize collaborative judging and mutation-driven analysis for general code-generation tasks. Building on these directions, our benchmark integrates real-world GitHub issue programs and loop-intensive SV-COMP cases, evaluates multiple prompting modes (zero-shot, few-shot, conversational, and collaborative), and systematically tracks verifier feedback, runtime, and validation effort through OpenJML logs. This design enables a comprehensive assessment of both effectiveness and efficiency in formal specification generation.

\section{Threats to Validity}
\label{sec:threats}
\subsection{Internal Validity}
Several factors could influence our findings. The wording and structure of prompts can significantly affect LLM responses. Even minor changes in phrasing or example selection, despite borrowing patterns from Xia et al.~\cite{xia2024automated}, may lead to different behaviors. Second, our evaluation benchmark includes programs drawn from SpecGenBench~\cite{specgen} and SV-COMP, both of which contain open-source samples that may have appeared in the pretraining data of proprietary LLMs. Because the training corpora of these models are not publicly disclosed, indirect exposure through related problems or patterns cannot be completely ruled out. To mitigate this risk, we manually compared the generated specifications against reference oracles and publicly available JML examples, finding no lexical or structural overlaps. In addition, our evaluation focuses on formal verification outcomes, as each specification is validated through OpenJML, so correctness is determined by logical consistency rather than textual similarity. These checks substantially reduce the likelihood that data leakage or memorization materially influenced the reported results. We acknowledge that a definitive leakage analysis would require membership inference or embedding-space similarity testing, which we plan to perform in future work to quantify residual exposure. Finally, our completeness measure is based on mutation testing. We employ Equivalent Mutant Suppression (EMS)~\cite{kushigian2024equivalent} to identify mutants that do not alter program semantics, but some equivalent variants may still slip through. This artifact can slightly overstate the completeness score. Nonetheless, since all methods in our comparison use the same mutation pipeline, the relative differences remain valid.

\subsection{External Validity}
Our evaluation relies on OpenJML for verifying JML-annotated programs. Like any deductive verifier~\cite{ahrendt2005key}, OpenJML cannot decide every proof obligation; there are correct specifications that it either rejects or marks \textit{unknown} because of underlying solver limitations~\cite{abdulla1996undecidable, mathur2020s}. Despite these verifier limitations, \name consistently produced valid, provable results for most test programs, including those borrowed from real-world GitHub issues. Our evaluation is limited to Java/JML verified with OpenJML; therefore, while \name is modular, our results do not yet demonstrate language-agnostic performance beyond this setting. We leave cross-language validation to future work.

\section{Conclusion and Future Work}
\label{sec:conclusion}
We introduced \name, a novel collaborative framework for verifiable specification generation that integrates dynamic LLM recommendation with verifier-guided prompt refinement. The system adjusts to the structure of the input program by selecting an LLM pair and a prompting strategy accordingly. This architecture allows \name to generate more accurate and verifiable specifications, even for complex program constructs. Our evaluation of 72 Java programs from benchmark suites and real-world bug reports shows that \name improves both verification success and completeness compared to previous tools, while significantly reducing evaluation time. To support adoption and future research, we release \name as an open-source framework, including a VS Code extension for interactive use and a public leaderboard for reproducible evaluations on our benchmark. Looking ahead, we plan to extend \name to other specification languages (e.g., ACSL, Viper), add lightweight static checks for early filtering, and explore on-the-fly prompt adaptation from verifier feedback. We also aim to scale to multi-module systems and concurrent APIs, revisit model selection as LLMs evolve, and broaden support to non-functional specifications (e.g., fairness, robustness).

\section{Data Availability}
\label{sec:datapackage}
The reproducibility package, evaluation results, the full benchmark, and leaderboard~\cite{AutoReSpecLeaderBoard} are available in an anonymous repository~\cite{AutoReSpecRepository} and we hope it serves as a useful resource for future research in this area.

\balance
\bibliographystyle{ACM-Reference-Format}
\bibliography{reference}



\end{document}